\documentclass[twocolumn,showpacs,preprintnumbers,amsmath,amssymb,prb,floatfix]{revtex4}

\usepackage{graphicx}
\usepackage{dcolumn}
\usepackage{bm}
\usepackage{epstopdf}    
 
\begin{document}
\bibliographystyle{prsty}

\title{Electronic Phases and Phase Separation in the Hubbard-Holstein Model of a Polar Interface}
\author{B. R. K. Nanda and S. Satpathy}
\affiliation{Department of Physics, University of Missouri, Columbia, Missouri 65211, USA  
}
\date{\today}

\begin{abstract}

From a mean-field solution of the  Hubbard-Holstein model, we show that a rich variety of different electronic phases can result at the interface between two polar materials such as   LaAlO$_3$/SrTiO$_3$. Depending on the strengths of the various competing interactions, viz., the electronic kinetic energy, electron-phonon interaction, Coulomb energy, and electronic screening strength, the electrons could (i) either be strongly confined to the interface forming a 2D metallic or an insulating phase, (ii) spread deeper into the bulk making a 3D phase, or (iii) become localized at individual sites forming a Jahn-Teller polaronic phase. In the polaronic phase, the Coulomb interaction could lead to unpaired electrons resulting in magnetic Kondo centers. Under appropriate conditions, electronic phase separation may also occur resulting in the coexistence of metallic and insulating regions at the interface. 

 \end{abstract}


\maketitle

\section{Introduction}
Transport measurements of the polar interfaces such as LaAlO$_3$/SrTiO$_3$
(LAO/STO) have revealed a rich variety of behavior\cite{huijben} such as a Kondo resistance minimum, metallic, insulating, and superconducting characters, as well as magnetism under varying experimental conditions. Superconductivity has also been observed\cite{reyren} below 200 mK and while the resistance minimum can be explained by either the 2D weak localization or the presence of Kondo centers, it has been argued that the observation of a negative magnetoresistance together with the logarithmic increase of resistivity with decreasing temperature suggests the presence of localized Kondo centers.\cite{brinkman} The resistance also follows the scaling laws for the Kosterlitz-Thouless transition in 2D,\cite{kosterlitz} which can be tuned by the carrier concentration induced by a gate voltage.\cite{caviglia} Yet another intriguing experimental discovery is the reversible switching between nanoscale conducting and insulating regions and patterning of conducting lines on the surface of the polar heterostructure using an atomic force microscopy probe.\cite{cen} 

The origin of the 2D electrons at the polar interfaces is becoming increasingly clear to be due to electronic reconstruction necessary to quench the polar catastrophe, i.e., the diverging electrostatic potential due to the lining up of the charged atomic planes must be mitigated by transferring electrons to the interface. However, the complex behavior of these electrons originating from the competition between the various interactions such as Coulomb and Jahn-Teller (JT) terms and external factors such as impurities and gate voltage that control the carrier density is less understood. 

In this paper, we study a model polar system using mean-field theory and examine the various phases of the electrons at the interface. We find that depending on the strength of the various competing interactions, the interfacial electrons show a rich variety of phases. Under appropriate circumstances, a phase separation between metallic and insulating phases is also possible.

%
\begin{figure}
\centering
\includegraphics[width=6.0cm]{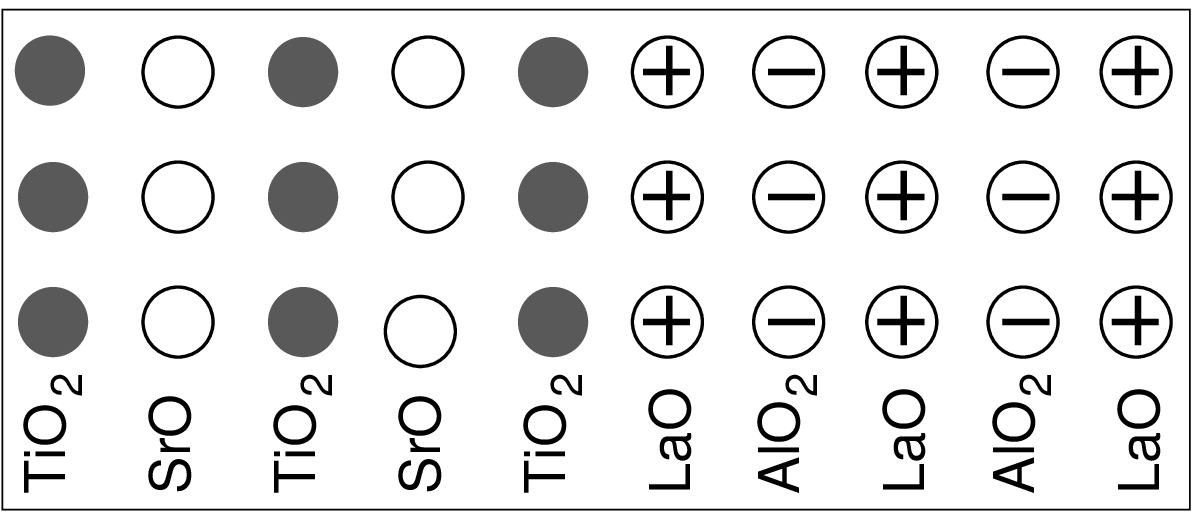}  \\
\vspace{5mm}
\includegraphics[width=7.0cm]{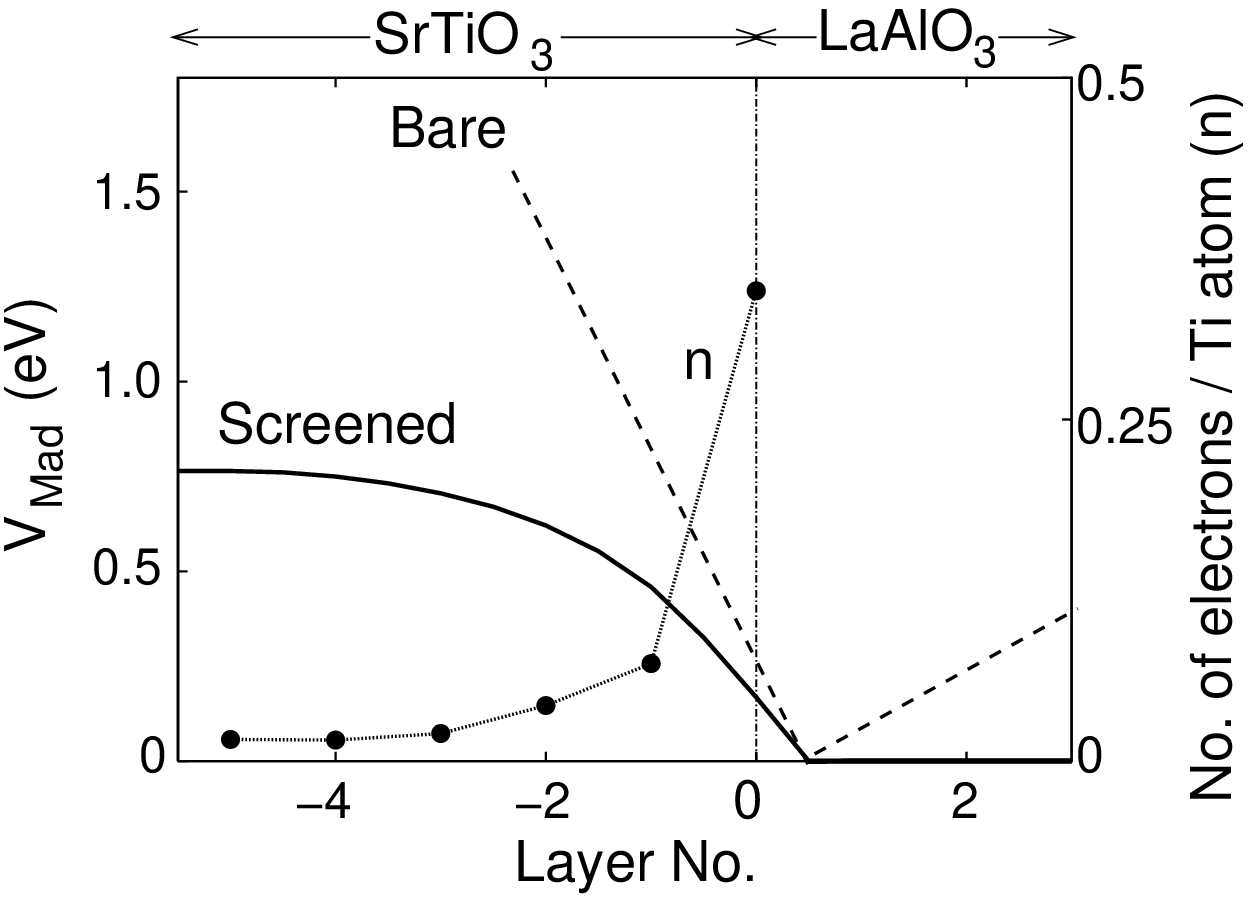}
\caption{
Model of a typical polar interface considered in the paper (Top). Here we are concerned with the $n$-type structure with TiO$_2$ and LaO forming the interfacial layers. Electrons migrate to the interface in response to the diverging Coulomb potential of the charged planes and interact via Coulomb and JT interactions, resulting in a rich variety of interfacial phases.
Bottom part shows the calculated bare and screened Coulomb potentials  and the accumulated electron density at the interface in response to the polar catastrophe (indicated by the bare potential, dashed line). Potentials shown are the computed Madelung potentials on the oxygen sites in the SrO or LaO layers. 
}
\label{potentialcharge}
\end{figure}
\section{Model and Method}
Before discussing the model Hamiltonian, we present a brief overview of the relevant electron states at the polar interfaces, taking the example of LAO/STO for the sake of concreteness.   Simple electron counting argument shows that half an electron per interfacial Ti atom are needed for the ideal interface to quench the divergence of the electrostatic potential. 
The electrons form a quasi-2D electron gas on the STO side, where they occupy the Ti(d) bands as indicated from 
density-functional calculations\cite{popovic, pentcheva1, janicka, lee, son, chen, park} as well as a variety of experiments\cite{huijben}.
 They furthermore become coupled to the lattice, as the Ti in SrTiO$_3$, originally in the d$^0$ state, becomes a d$^1$ JT ion.   Crystal field splits the Ti t$_{2g}$ states  below e$_g$\cite{STO-bulk}, which then interact via the JT interaction with several octahedral stretching and shear modes, leading to a complex distortion of the octahedra at the interface.\cite{Kelly}
In the present work,  we restrict to the simplified Holstein model of the electron-lattice interaction, which describes a single orbital degree of freedom. The model captures the essential physics of the competing interactions including the polaron formation, although effects such as orbital ordering are obviously not contained in the model. 

The  Hubbard-Holstein model adopted here describes the motion of the electrons on the Ti sites with the other ionic charges serving as a Coulombic background. The Hamiltonian is
\begin{equation}
{\cal H} = {\cal H}_{Mad} + {\cal H}_{e}  + {\cal H}_{e-p} ,
\label{htot}
\end{equation}
where the first term describes the Madelung energy
\begin{equation}
{\cal H}_{Mad} = \frac{1}{\epsilon} \sum_{ij}^\prime  M_{ij} q_i q_j,
\label{hpolar}
\end{equation}
$M$ being the Madelung matrix, which may be calculated using the standard Ewald's method, $i$ and $j$ are the site indices, $q_i$ is the total of the ionic charge $q_i^{\prime}$ and the electronic charge $n_i$ on the $i$-th site ($q_i=q_i^{\prime} +n_i$), $\epsilon$ is the static dielectric constant, and the prime on the summation excludes same site terms.  Note that while the site summations run over all atoms, the electronic charge $n_i$ is non-zero on the Ti sites only. Without the electronic charges the Madelung energy would diverge as the potential grows unrestricted away from the interface (polar catastrophe). The various ions are passive, merely providing a Coulomb potential background for the Ti electrons to move.

The second term  represents the electronic kinetic energy and the electron-electron interaction 
\begin{equation}
{\cal H}_{e} = \sum_{\langle ij \rangle \sigma} -t \  c_{i\sigma}^\dagger c_{j\sigma} + H. c. +  U \sum_i  n_{i\uparrow} n_{i\downarrow},
\label{hke}
\end{equation}
where $c_{i\sigma}$, $c^{\dagger}_{i\sigma}$ are  the field operators, $\sigma$ is the spin index, and $\langle ij \rangle$ indicates summation over distinct nearest-neighbor pairs. An on-site Coulomb repulsion (U) is retained in the electron-electron interaction term.
%
The final term in the Hamiltonian is the electron-lattice interaction within the Holstein model 
\begin{equation}
{\cal H}_{e-p} = \sum_i (\frac{1}{2}K Q_i^2 - gQ_i n_i),
\label{hjt}
\end{equation}
where $Q$ is the lattice distortion at a lattice site, K and g are the stiffness constant and the electron-lattice coupling strength, respectively, and $n_i$ is the electron occupancy of the $i^{th}$ Ti atom. In a degenerate orbital model, $g$ is the Jahn-Teller coupling strength, while in the Holstein model, it represents the coupling of the lattice to the energy of the single orbital present in the model. For a strong enough $g$, an electron becomes localized to the lattice site forming a Jahn-Teller (or Holstein) polaron. For a given K and g, the magnitude of Q can be minimized to obtain the minimum energy which yields the site distortion $Q_i$ = $gn_i/K$, so that the  elastic energy in the system becomes ${E}_{elastic} =  \frac{1}{2}\sum_{i}g^2n{_i}^2/K$.
 %
%

The Hamiltonian is self-consistently solved within the mean-field Hartree approximation : $n_{i\uparrow}n_{i\downarrow} = n_{i\uparrow}\langle n_{i\downarrow}\rangle + \langle n_{i\uparrow}\rangle n_{i\downarrow} - \langle n_{i\uparrow}\rangle\langle n_{i \downarrow}\rangle$. In the calculations we used a supercell geometry of eleven layers of STO and six layers of LAO, with the unit cell doubled along the interface in order to study charge disproportionation and polaron formation at the interface. The supercell geometry contained two identical $n$-type interfaces of the type shown in Fig. \ref{potentialcharge}.
The ground state energy is given by
\begin{eqnarray}
E_{tot}&=&\frac{1}{N_{\kappa}}\sum_{\kappa,\sigma}^{occ}\varepsilon_{\kappa\sigma}+\frac{1}{2}\sum_{i}g^2n{_i}^2/K \\ \nonumber
&+& \frac{1}{\epsilon}\sum_{ij}^{\prime}M_{ij} q_i^{\prime}q_j^{\prime}  
- \frac{1}{\epsilon}\sum_{ij}^{\prime}M_{ij}  n_i n_j  
-\sum_i U\langle n_{i\uparrow} \rangle \langle n_{i\downarrow} \rangle,
\label{etot}
\end{eqnarray}
where the last two terms are the double-counting terms and $N_{\kappa}$ is the number of $\kappa$ points used in the Brillouin zone summation.

Typical parameters for the Hamiltonian are: electron hopping $-t \approx  - 0.15$ eV, $ K \approx 5-10$ eV / \AA$^2$, $g \approx $ 2 eV/\AA, $\lambda = g^2/2K \approx 0.2$ eV, and $U \approx 2 - 5$ eV. We have used $U = 2$ eV in our calculations; the essential effect of the Coulomb $U$ in our model is to prevent double occupancy of the lattice sites due to the Holstein interaction parameter $\lambda$. This being the case, a larger $U$ does not make any essential change to our result.
The dielectric constant $\epsilon$ can vary over a relatively wide  range.
For example, for bulk SrTiO$_3$, it changes between 25000 and 300 as temperature changes from 4K to 300K\cite{sakudo} and it decreases sharply with an electric field.\cite{etsuro, copie} The internal electric field at the LAO/STO interface, for example, is somewhere between 1 - 10 mV/{\AA}\cite{copie, chen1, bristowe, pentcheva2, ramesh},  which leads to $\epsilon \sim 10 - 100$ following the non-linear expression for $\epsilon$ as a function of the electric field\cite{copie}.  Our earlier density-functional calculations\cite{popovic-relaxed} showed an average dielectric constant at the LAO/STO interface to be $\epsilon \sim 68$, when lattice relaxation is taken into account, which is consistent with   the above values.

\section{Results and Discussion}

Fig. \ref{potentialcharge} shows the typical electron distribution at the interface as well as the bare Coulomb potential coming from the ionic charges 
(La$^{+3}$, Al$^{+3}$, Sr$^{+2}$, Ti$^{+4}$, and O$^{-2}$), which diverges away from the interface (polar catastrophe) and the screened potential produced by the transfer of the electrons to the interface. Our supercell consisted of two identical n-type TiO$_2$/LaO interface, so that exactly half an electron per interface Ti atom is available to satisfy the polar catastrophe. As seen from the figure, there is no divergence in the screened potential, while there is a potential shift due to the accumulated dipole moment at the interface.

\begin{figure}
\centering
\includegraphics[width=7.0cm]{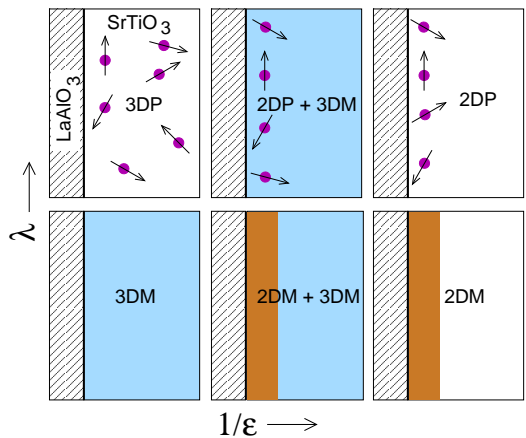} 
\caption{(Color online) Electronic phases at the polar interface (schematic). In the 2D phase,  electrons are confined very close to the interface (one or two monolayers), while in the 3D phase, they spread out deeper into the STO bulk. Both cases can be metallic (M) or polaronic (P). In the polaronic phase, electrons are localized to individual sites due to a strong electron-lattice interaction, which results in an insulating phase leading to unpaired electrons on individual Ti sites (possible Kondo centers), if the on-site Coulomb U is strong enough.}
\label{summaryresult}
\end{figure}

{\it Competing phases} -- Our model shows a variety of different phases depending on the 
strength of the competing interactions, which are illustrated in Fig. \ref{summaryresult}. If the 
 dielectric constant $\epsilon$ is large, the Coulomb potential of the ions (Madelung potential)
 that confines the electrons to the interface becomes weaker, thereby spreading out the electrons deeper  into the SrTiO$_3$ side, resulting in a more 3D like electron gas. The opposite happens if $\epsilon$ is small, so that the electrons are confined to just one or two monolayers at the interface, leading to a 2D electron gas. In either case, a polaronic phase (2DP or 3DP)
 can form, if the electron-lattice coupling $\lambda$ is strong enough, so that electrons are confined to individual sites, giving up the kinetic energy they would have gained by delocalization. An on-site Coulomb interaction on the Ti site prevents the occupation of two such electrons at the same site, so that
 unpaired electrons can form magnetic centers.

\begin{figure}
\centering
\includegraphics[width=6.5cm]{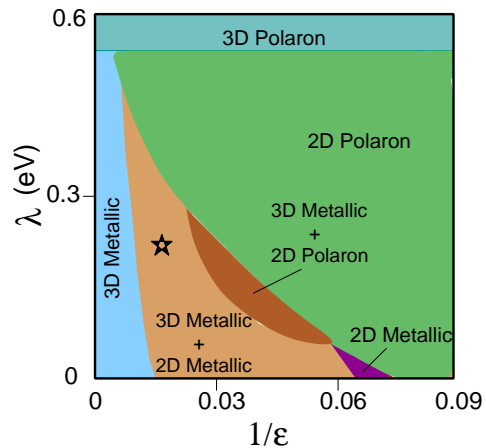} 
\caption{ (Color online) Phase diagram for the Hubbard-Holstein model with  half an electron per interface unit cell. The star indicates the approximate parameters 
for the LAO/STO interface, which is near several phase boundaries, leading to the possibility of phase separation as discussed in the text. 
}
\label{phaseu3}
\end{figure}

\begin{figure}
\centering
\includegraphics[width=8.5cm]{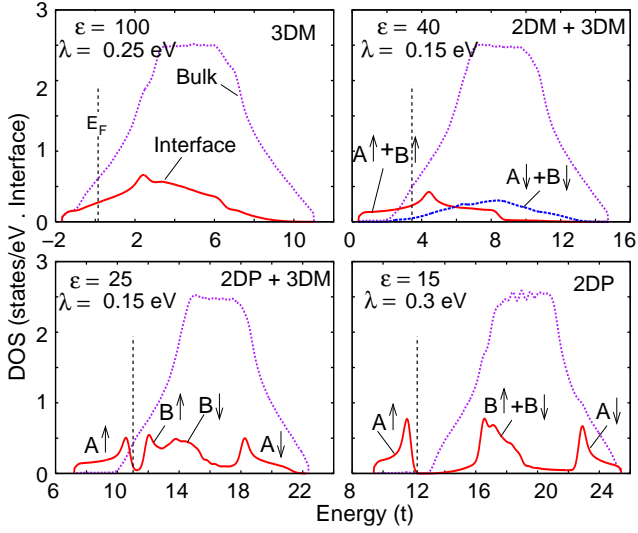} 
\caption{(Color online) Densities-of-states for the interfacial layer and the remaining layers (bulk) illustrating the characteristics of the various phases.  Electron states are occupied up to the Fermi energy $E_F$. The letters A and B indicate the two inequivalent Ti sites in the interfacial TiO$_2$ plane and arrows indicate the two spin states. In the case of the 2DP phase for example (bottom right), all electrons are confined to the interface layer and occupy the A sites only, one unpaired electron per site, while in the 3DM phase (top left), the electrons spread out into both the interface sites and sites deeper in the bulk.
}
\label{ldos}
\end{figure}

The Hubbard-Holstein model captures the essence of the competition between electrostatic and electron-lattice coupling and a  phase diagram as a function of $\epsilon$ and $\lambda$ obtained from this model is shown in Fig. \ref{phaseu3}. For the typical parameters for the LAO/STO interface,  $\lambda \approx 0.2 $ eV and $\epsilon \approx 70$,  the model shows a combined 3DM + 2DM behavior shown by a star mark in the figure. In this phase, some electrons are confined to the interfacial layers while some are spread deeper into the bulk as indicated from the projected densities  of states (DOS) in Fig. \ref{ldos}. This phase is consistent with the density-functional calculations for the interface as well as with the experimental studies with very low oxygen pressure which shows a mixture of 2D and 3D metallic behavior\cite{copie, seimon}.

The typical one-electron DOS for the various phases is shown in  Fig. \ref{ldos}. 
For large $\epsilon$ (top left of Fig. \ref{ldos}), the confining potential at the interface due to the Madelung term is weak, so that the electrons reside both in the interfacial and the bulk layers to form a 3D metallic phase. With decreasing $\epsilon$, the confining potential becomes stronger, leading to an increasing confinement of the electrons to the interface layer forming either a 2DM + 3DM phase or a completely 2DM phase. If the electron-lattice  coupling parameter $\lambda$ is strong enough, the
2D electrons become localized to individual Ti sites  forming a 2D polaron phase (bottom right of Fig. \ref{ldos}).

{\it Charge disproportionation and polaron formation} -- In order to examine the instability towards polaron formation with the increase of the electron-lattice coupling $\lambda$, we have performed a constrained optimization with respect to the charge disproportionation using the standard Lagrange multiplier technique. The constrained energy is written as
\begin{equation}
E(\delta) = E (\{ n_i \}) + \eta (n_A-n_B -\delta),
\end {equation} 
where the charge disproportionation on the two inequivalent interface sites, $ n_A - n_B$, is constrained to the value $\delta$ and $\eta$ is a Lagrange multiplier. The Lagrange multiplier may be written as the partial derivative of the constrained energy
$\partial E / \partial \delta = - \eta$, so that the constrained energy may be obtained by integrating the Lagrange multiplier
 \begin{equation}
E(\delta) = -\int_0^\delta \eta(\delta) \ d\delta.
\label{Edelta} 
\end {equation} 

\begin{figure}
\centering
\includegraphics[width=5.5cm]{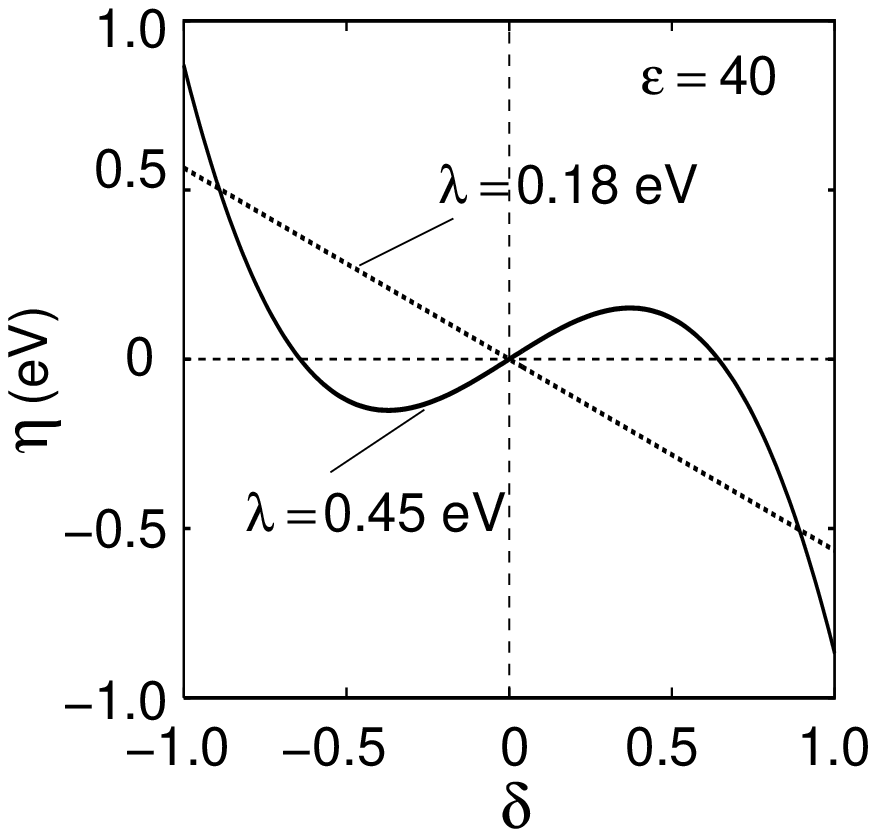} 
\includegraphics[width=5.5cm]{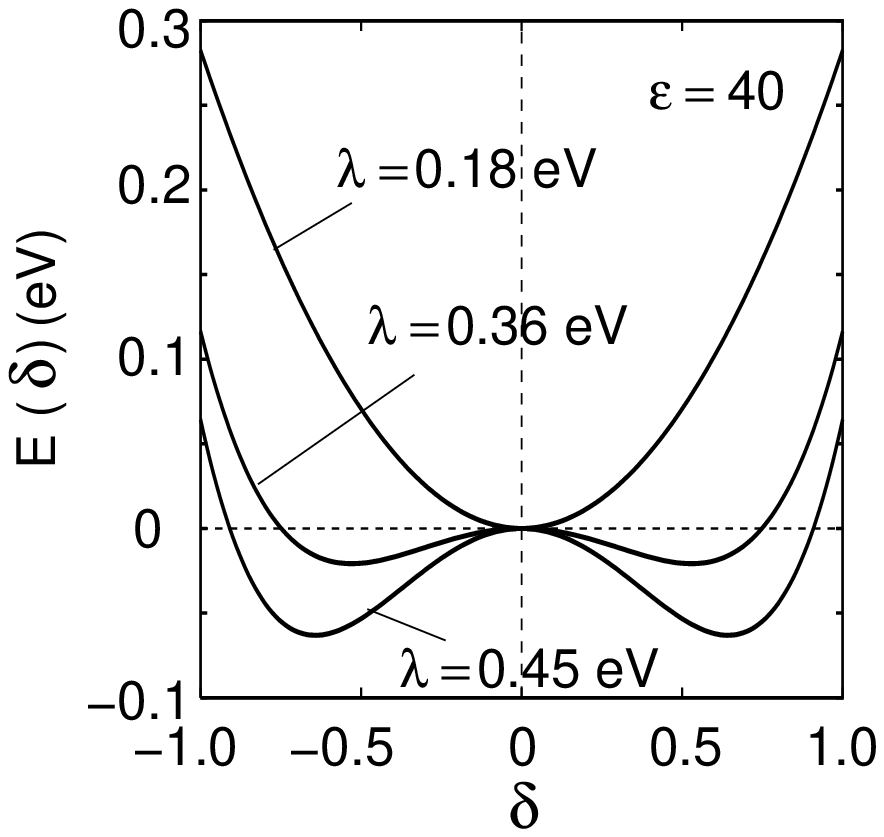} 
\caption{Force of constraint $\eta$ as a function of the constrained disproportionation 
parameter $\delta$ (top) and the energy for several values of the electron-lattice coupling strength $\lambda$ (bottom) as obtained from the expression Eq. (\ref{Edelta}). Charge disproportionation occurs beyond a critical value of $\lambda$.
} 
\label{constrained}
\end{figure}

\begin{figure}
\centering
\includegraphics[width=6.0cm]{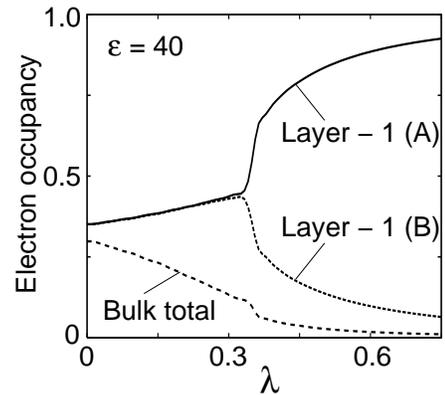} 
\caption{Electronic charge in the A and B interfacial sites and the rest of the sites (bulk)
corresponding to the minima of Fig. \ref{constrained} for various values of $\lambda$.
Beyond a critical value of $\lambda$, a 2DP phase is obtained, where the A sites accommodate an electron, leaving the B sites more or less empty.
}
\label{2DP}
\end{figure}

The results of the constrained-$\delta$ calculations are shown in Figs.  \ref{constrained} and \ref{2DP}.
For small $\lambda$, the minimum energy occurs at $\delta$ = 0, which indicates that all interfacial Ti sites are uniformly occupied. However as $\lambda$ increases, there is an instability as the kinetic energy lost by localizing the electron at an individual site is overcome by the energy gained by the formation of the Holstein polaron. A large enough Coulomb repulsion $U$ prevents the accumulation of two electrons at the same site.

%

{\it Electronic phase separation } -- The above analysis shows  charge disproportionation between the two interfacial sites, leading to the A sites which are electron rich and B sites which are electron poor. Such electron-rich sites may congregate together resulting in electron-rich regions and electron-poor regions. In fact, such mixed phases are ubiquitous and have been observed in many complex oxides including 
manganites\cite{dagotto,tokuraphase} and cobaltates\cite{phelanphase} and the presence of such inhomogeneous phases
at the polar interfaces has been suggested to explain transport measurements. 

Such inhomogeneous phases can in principle be studied by considering very large unit cells, but since the size of the 
individual phase regions could be several nanometers, the size of such supercells makes such study, either from density-functional or from the model Hamiltonian studies, computationally prohibitive. To address this issue, we study the energy of the individual homogeneous phases with different electron densities. This however makes the individual phases charge non-neutral, making their Coulomb energy diverge. But, the phase separated system is globally neutral so that the divergences of the individual phases cancel out and one need not consider these divergences explicitly. The Coulomb interaction within and between the phases means that the phases would be intermixed in the nanometer scale leading to the so-called Coulomb frustrated nanoscale phase separation.  What we study here is the tendency of the material to phase separate and not the pattern of the mixed phase, which is more complicated requiring a detail balance between the Coulomb energy, surface energy of the phase boundaries, in addition to the energetics of the individual phases.

To study the electronic phase separation, we varied the electron density $n$ of the homogeneous phase and calculated the total energy $E(n)$, leaving aside the divergent Coulomb energy of the charged phase, which would cancel for the global system as argued above. As a result, the expression for the   Madelung energy Eq. \ref{hpolar} becomes   
\begin{equation}
{\cal H}_{Mad} = \frac{1}{\epsilon} \sum_{ij}^\prime  M_{ij} (q_i q_j - \delta n_i \delta n_j), 
\label{hpolar2}
\end{equation}
where $M$ is the standard Madelung matrix, $q_i$ is the total site charge including ionic and electronic of the non-neutral system, and $\delta n_i = (n_i - n_i^0)$ is the excess charge at the i-th site, $n_i$ and  $n_i^0$ being the electronic charge for the charged phase and the neutral phase, respectively. The  above expression includes all Madelung energy except for the Coulomb energy of the excess charge, which would diverge with increasing size of the domain.
 
\begin{figure}
\centering 
\includegraphics[width=6.0cm]{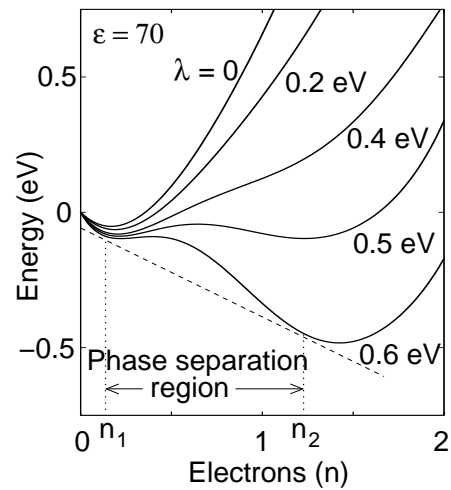} 
\caption{Energy as a function of electron concentration $n$ of the homogeneous phase indicating the existence of phase separation if the electron-lattice coupling $\lambda$ exceeds a critical value.
}
\label{phasesep}
\end{figure}

\begin{figure}
\centering
\includegraphics[width=6.0cm]{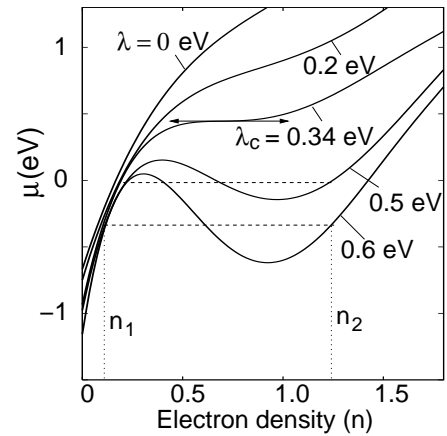} 
\caption{Chemical potential as a function of electron density at T = 0 and for parameters
corresponding to Fig. \ref{phasesep}. Phase separation occurs beyond the critical electron-lattice coupling of $ \lambda_c$.
}
\label{chempot}
\end{figure}

\begin{figure}
\centering
\includegraphics[width=6.5cm]{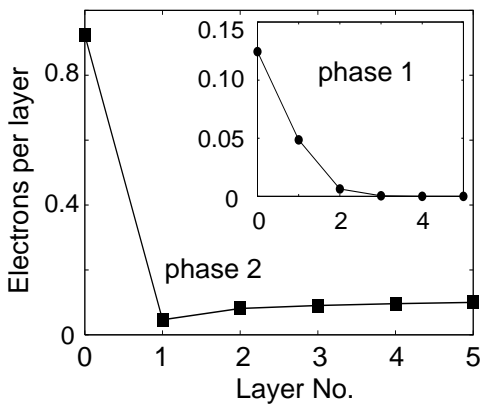} 
\caption{Electron distribution in various layers indicating the electron-rich polaronic phase (phase 2) and 
the electron-deficient  metallic phase (phase 1) for the case of $\lambda$ = 0.6 eV in Fig. \ref{phasesep}.
}
\label{phasen}
\end{figure}

 In Fig. \ref{phasesep} we have plotted the total energy $E(n)$ with respect to the electron concentration $n$. Note that the average concentration for the global system is $n = 0.5$ per interface Ti atom. For the stability of a homogeneous phase, $E(n)$ must be convex and the figure indicates that if the electron-lattice coupling exceeds a critical value $\lambda > \lambda_c \approx 0.34$, then the system phase separates into an electron-rich phase of density $n_1$ and an electron-poor phase of density $n_2$.  A similar conclusion was obtained by Kumar {\it et al.}\cite{kumar} from the study of a finite lattice model in two dimensions. The chemical potential $\mu = \partial E/ \partial n$ plotted in Fig. \ref{chempot} shows the phase separated region obtained from the Maxwell construction.

To understand the nature of the two phases, we have plotted in Fig. \ref{phasen}  the layer distribution of the electronic charge for the electron-rich and the electron-deficient phases. In the electron-rich phase $n = n_2 \approx 1$ and the electrons occupy the interface Ti sites with occupancy nearly equal to one, while in the electron-poor phase $n = n_1 $, which is nearly zero. These two phases will intermix with the average density of $n = 0.5$, which is sufficient to quench the polar catastrophe. Phase 1 may be characterized as a low density 3D metallic region, while phase 2 is a high-density 2D insulating region. Thus the interface region may be partly insulating and partly conducting leading to a lower number of electrons participating in transport  as has been seen for example in the LAO/STO interface.
	
\begin{figure}
\centering
\includegraphics[width=6.0cm]{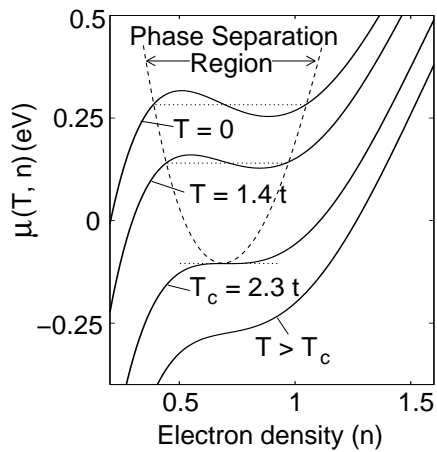} 
\caption{Chemical potential as a function of temperature and electron density for the 
case corresponding to $\lambda = 0.5 $ eV in Fig. \ref{chempot}. Temperature T is in units of $k_B^{-1}$.
}
\label{muT}
\end{figure}

{\it Estimation of $T_c$ for phase separation} -- The tendency towards phase separation diminishes with increasing temperature and disappears beyond a critical value $T_c$. At a finite temperature, one needs to calculate the free energy and the chemical potential $\mu(T,n)$ as a function of temperature and density. For $T=0$, the free energy is $f = \varepsilon$ and $\mu = \partial (n\varepsilon)/\partial n$. For finite $T$, we compute  $f$ and $\mu$ using a model of non-interacting fermions and assuming that only the electronic energy changes with temperature.  An estimate of $T_c$ may be obtained by using the low temperature expansion of the free energy
\begin{equation}
F(T)= F(0) - \alpha T^2 / 2,
\end{equation}
where $\alpha$ is the coefficient of the Sommerfeld specific heat. Furthermore, we approximate the one-particle density of states by a semi-circular form
\begin{equation}
D(E) = 2\pi^{-1} B ^{-2} \sqrt {B^2 - E^2},
\label{semicircular}
\end{equation}
$B \approx 1 $ eV being half of the band width. The chemical potential is then given by
\begin{equation}
\mu (T, n) = \mu(0) +\frac{\pi k_B^2 T^2}{6} \frac{E_F}{B^2-E_F^2},
\end{equation}
where the Fermi energy $E_F$ is a function of the electron density $n$. The results are plotted in Fig. \ref{muT}, which shows the transition from a mixed phase to a single phase beyond the critical temperature $T_c$, whose exact magnitude depends on the relative strengths of the various interactions.   

{\it Domain size} -- So far we have studied the instability of the interface to phase separation into two charged phases. However, as discussed by many authors,\cite{Lorenzana1, Lorenzana2, Kugel} such phase separation is frustrated due to the long-range Coulomb interaction, leading to a strong nanoscale intermixing between the single-phase domains in order to minimize the Coulomb energy. The size of the domains in the present case may be estimated by minimizing the energy of the electron-rich phase. Taking the electron-rich phase to have a two-dimensional disk structure, the energy of N electrons occupying a single site each in a circle of radius R may be evaluated ($\pi R^2 = N a^2$, where $a$ is the lattice constant). 

The energy gain due to the Holstein (or JT) polaron formation for each electron is $-\lambda $, while the confinement energy of the electron decreases with the domain size 
going as $\gamma \ \hbar^2 \pi^2(2 m R^2)^{-1}$, where $\gamma = \nu^2/\pi^2 \approx 0.58$, $\nu$ being the zero of the Bessel function $J_0(\nu) =0$, and the Coulomb energy increases as $2 \sigma^2 (3\epsilon)^{-1}R^3$, $\sigma$ being the surface charge density of the disk. Putting these together, the energy per particle $E$ may be expressed as
\begin{equation}
E = -\lambda  +  \gamma \  \frac{\hbar^2 \pi^3}{2 m a^2}  N^{-1} + \frac{2 e^2}{3 \pi^{3/2}\epsilon a}  N^{1/2},
\end{equation}
the minimization of which yields the optimal domain size 
$ N = ( 3 \gamma/ (8 \pi))^{2/3} \pi^3 (\epsilon_r  (m^* a_l)^{-1})^{2/3}$. Taking the relative dielectric constant at the STO interface $\epsilon_r \approx 70$, band effective mass
$m^* \approx 10$, and lattice constant in units of the Bohr radius $a_l \approx 8$, the domain size is estimated to be $N \approx 5$ or $2R \approx 10$ \AA. This provides a rough estimate of the domain size, which would intermix to minimize the long-range Coulomb energy between the electron-rich and the electron-poor domains.

\section{Summary}
To summarize, we studied the Holstein-Hubbard model for a polar interface such as LAO/STO in the mean-field approximation, which showed the existence of a rich variety of electronic phases. The phases could be  metallic or insulating  and two or three dimensional depending on the strengths of the competing interactions. Our study also suggests the formation of phase separation below a critical temperature, which makes the transport behavior in these materials quite rich. The phase separation scenario with mixed metallic and insulating phase offers a possible explanation for the observation of a lower density 2DEG at the interface than what is necessary to satisfy the polar catastrophe. The phase separation is however frustrated due to the long-range nature of the Coulomb interaction leading to the formation of  domains with sizes on the scale of several nanometers as has been observed in the complex oxides such as the manganites. 

We thank Shi-Jie Chen for stimulating discussions. This work was supported by the U. S. Department of Energy through Grant No. DE-FG02-00ER45818.


\end{document}